\documentclass[aps,amsmath,amssymb,prl,10pt,showpacs,twocolumn,letterpaper,superscriptaddress,showonlyrefs]{revtex4-2}
\usepackage{graphicx,color,tcolorbox,txfonts}
\usepackage{bm}
\usepackage{mathrsfs}
\usepackage{comment}
\usepackage{mathtools}
\usepackage{hyperref}
\usepackage{epstopdf}
\usepackage{letltxmacro}
\newcommand{\al}{\alpha}
\newcommand{\be}{\beta}
\newcommand{\g}{\gamma}
\newcommand{\de}{\delta}

\newcommand{\io}{{\rm i}}

\newcommand{\p}{\pi}

\newcommand{\s}{\sigma}
\newcommand{\y}{\upsilon}
\newcommand{\f}{\phi}

\newcommand{\w}{\omega}
\newcommand{\W}{\Omega}
\newcommand{\De}{\Delta}
\newcommand{\G}{\Gamma}

\newcommand{\ta}{\tau}

\renewcommand{\y}{\psi}

\newcommand{\pd}{\partial}

\newcommand{\round}[1]{\left( #1 \right)}
\renewcommand{\square}[1]{\left[ #1 \right]}
\newcommand{\curly}[1]{\left\{#1\right\}}

\newcommand{\sang}[1]{\langle #1 \rangle}

\newcommand{\beq}{\begin{equation}}
\newcommand{\eeq}{\end{equation}}
\newcommand{\Beq}{\begin{eqnarray}}
\newcommand{\Eeq}{\end{eqnarray}}
\newcommand{\bml}{\begin{multline}}

\newcommand{\bea}{\begin{align}}
\newcommand{\ena}{\end{align}}
\newcommand{\bsp}{\begin{split}}
\newcommand{\esp}{\end{split}}

\newcommand{\br}{{\boldsymbol r}}
\newcommand{\bM}{{\boldsymbol M}}

\newcommand{\bs}{{\boldsymbol s}}
\newcommand{\bp}{{\boldsymbol p}}

\newcommand{\bk}{{\boldsymbol k}}

\newcommand{\bq}{{\boldsymbol q}}

\newcommand{\bQ}{{\boldsymbol Q}}
\newcommand{\bB}{{\boldsymbol B}}

\newcommand{\cJ}{\mathcal{J}}

\newcommand{\bnab}{\boldsymbol{\nabla}}

\newcommand{\ve}{\varepsilon}

\newcommand{\bzero}{\boldsymbol{0}}

\newcommand{\bsig}{\boldsymbol{\sigma}}

\DeclareMathAlphabet{\mathpzc}{OT1}{pzc}{m}{it}

\begin{document}
\title{Conductivity Enhancement in a Diffusive Fermi Liquid due to Bose-Einstein Condensation of Magnons}
\author{Joshua Aftergood}
\affiliation{Department of Physics, Queens College of the City University of New York, Queens, NY 11367, USA}
\affiliation{Physics Doctoral Program, The Graduate Center of the City University of New York, New York, NY 10016, USA}
\affiliation{Department of Physics and Astronomy, Iowa State University, Ames, IA 50011, USA}
\author{So Takei}
\affiliation{Department of Physics, Queens College of the City University of New York, Queens, NY 11367, USA}
\affiliation{Physics Doctoral Program, The Graduate Center of the City University of New York, New York, NY 10016, USA}
\date{\today}

\begin{abstract}
We theoretically study the conductivity of a disordered 2D metal when it is coupled to ferromagnetic magnons with a quadratic spectrum and a gap $\De$. In the diffusive limit, a combination of disorder and magnon-mediated electron interaction leads to a sharp metallic correction to the Drude conductivity as the magnons approach criticality, i.e., $\De\rightarrow0$. The correction is non-singular and is distinctively weaker than, for example, the log-squared correction obtained when disordered electrons couple to diffusive spin fluctuations near a Hertz-Millis transition. The possibility of verifying this prediction in an $S=1/2$ easy-plane ferromagnetic insulator K$_2$CuF$_4$ under an external magnetic field is proposed. Our results show that the onset of a magnon BEC in an insulator can be detected via electrical transport measurements on the proximate metal.

\end{abstract}
\maketitle

In the low-temperature normal state of a pure metal, electron-electron interactions manifest themselves only in the renormalization of the electron spectral parameters~\cite{nozieresBOOK99}. In the presence of disorder, however, electrons propagate diffusively at distances longer than the mean-free path $\ell$, resulting in stronger interactions. These disorder-enhanced interactions, together with reduced dimensionality, lead to singularities in various thermodynamic and transport quantities~\cite{altshulerBOOK85}. Singularities in transport can also arise through the coupling of electrons to other dynamical degrees of freedom~\cite{belitzPRL00}. Several works have studied the conductivity of disordered Fermi liquids tuned close to magnetic quantum critical points. Near these critical points, disorder generates a $\ln^2T$ correction to the Drude conductivity~\cite{kimPRB03,paulPRL05,paulPRB08}, suggesting that diffusive critical spin fluctuations can also strongly enhance impurity scattering at low energies. Few works since then have studied the feedback of critical spin fluctuations on disordered electrons in the form of quantum conductivity corrections~\footnote{Non-Fermi liquid temperature scaling has been predicted in disordered two-dimensional metals near an antiferromagnet quantum critical point as well~\cite{roschPRL99,syzranovPRL12}. This work, however, will focus exclusively on critical behavior with a divergent $q=0$ susceptibility.}. It is therefore interesting to explore such corrections arising in the vicinity of a wider variety of critical points. 

\begin{figure}
\includegraphics[width=0.8\linewidth]{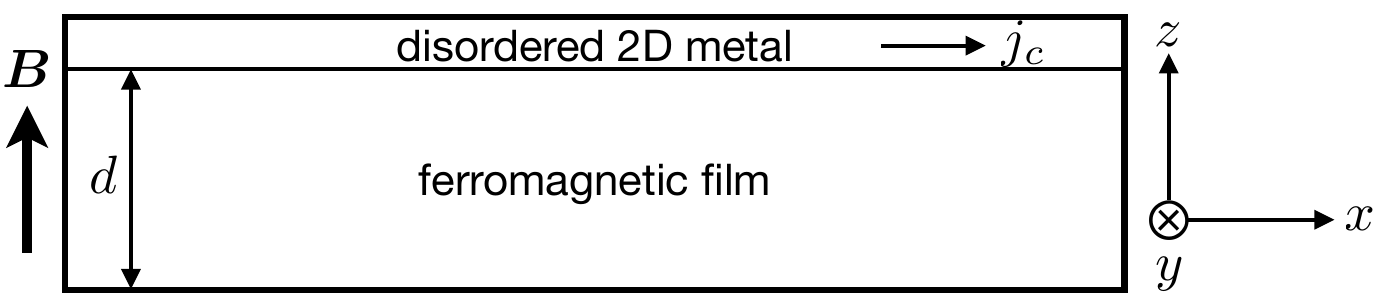}
\caption{A depiction of the bilayer system. The metal and the ferromagnetic insulator couple via the exchange coupling $\cJ$, and $j_c$ represents the charge current flowing in the metal along the $x$ axis. A magnetic field is applied along the $z$ axis.}
\label{fig1}
\end{figure}

To access a wider array of materials and therefore critical phenomena, we may consider a {\em bilayer} system affixing a ferromagnetic insulator (FI) to a disordered conductor (see Fig.~\ref{fig1}). When the FI approaches a quantum critical point with a divergent $(\bq,\W)=(\bzero,0)$ susceptibility, e.g., the equilibrium Bose-Einstein condensation (BEC) of ferromagnetic magnons, critical spin fluctuations injected into the metal layer can drive conductivity corrections. However, the presence of an exchange field due to the FI magnetization results in Zeeman splitting in the electron sector, introducing a gap to the diffusive, transverse spin density fluctuations in the metal. This gap leads to a detuning between the critical magnon mode of the FI and the diffusive spin density fluctuations in the metal, possibly weakening the influence the critical magnons have on the transport in the adjacent metal. A quantitative understanding of the effects of this detuning on the conductivity correction remains an open problem but may have relevance to the field of spintronics, where magnetoresistance phenomena in metal-FI bilayers are routinely studied. This understanding may also uncover an intriguing possibility of probing the onset of magnon BECs in FIs using charge transport measurements.

In this Letter, we theoretically address this problem by studying the conductivity of a disordered 2D metal when it is exchanged-coupled to a FI as depicted in Fig.~\ref{fig1}. We monitor the conductivity of the metal as the spin-1 magnons undergo BEC~\cite{nikuniPRL00,riceSCI02,giamarchiNATP08}. We find that this BEC transition leads to a sharp enhancement in the conductivity as $\De\rightarrow0$, where $\De$ is the distance to the BEC critical point. However, this enhancement is non-singular and is therefore distinct from the singular logarithmic corrections studied in the past~\cite{kimPRB03,paulPRL05,paulPRB08}. That said, this enhancement can be of order $10{\rm m\W}$ for a certain FI at $\De\approx0$ and should be detectable. A corollary of this finding is that a conductivity measurement on the adjacent metal at a fixed bilayer temperature can probe the BEC transition in the magnetic insulator through the detection of this sharp conductivity enhancement. While magnon BECs in quantum magnets are typically probed using, e.g., magnetic susceptibility, specific heat, and magnetocaloric measurements~\cite{zapfRMP14}, this work proposes an alternative probe of the onset of a magnon BEC based on charge transport measurements. 

\textit{Model}:~Let us consider a 2D metal deposited atop a FI of thickness $d$ with the interface held in the $xy$ plane (see Fig.~\ref{fig1}). The metal layer is modeled as a standard disordered electron gas with the Hamiltonian
\beq
H_e = \int d^2\br\,\y^\dag_{\br\s}\round{-\frac{\hbar^2\boldsymbol{\nabla}^2}{2m} - \mu+\frac{\s\hbar\w_{Z}}{2}+u(\br)}\y_{\br\s}\label{h0}\,,
\eeq
where summations over repeated spin indices are implied. Here, $\y_{\br\s}$ is the electron field operator, $\mu$ is the chemical potential, and $u(\br)=u\sum_{i}\de(\br-\br_i)$ is the short-ranged $s$-wave impurity potential; $\hbar\w_Z$ is the total electron Zeeman energy arising from the external field $\bB$ and the FI magnetization, both of which are parallel to the $z$ axis. 

A model system for studying magnon BEC is a FI with an easy-plane magnetic anisotropy and an external field applied normal to the easy-plane. For an $xy$ easy-plane, the ferromagnetic Hamiltonian may be written as 
\begin{multline}
\label{hf}
H_F=\int_{-d}^0dz\int d^2\br\,\bigg[\frac{A}{2s^2}\round{\bnab\bs(\br,z)}^2\\
+\frac{K}{2s^2}s^2_z(\br,z)+\hbar\g Bs_z(\br,z)\bigg]\,,
\end{multline}
where $A$ and $K>0$ parametrize the exchange stiffness and the anisotropy, respectively, $\g$ is the gyromagnetic ratio, $\bs(\br,z)$ is the local spin density, and $s$ is the saturated spin density. 

\begin{figure}
\includegraphics[width=0.8\linewidth]{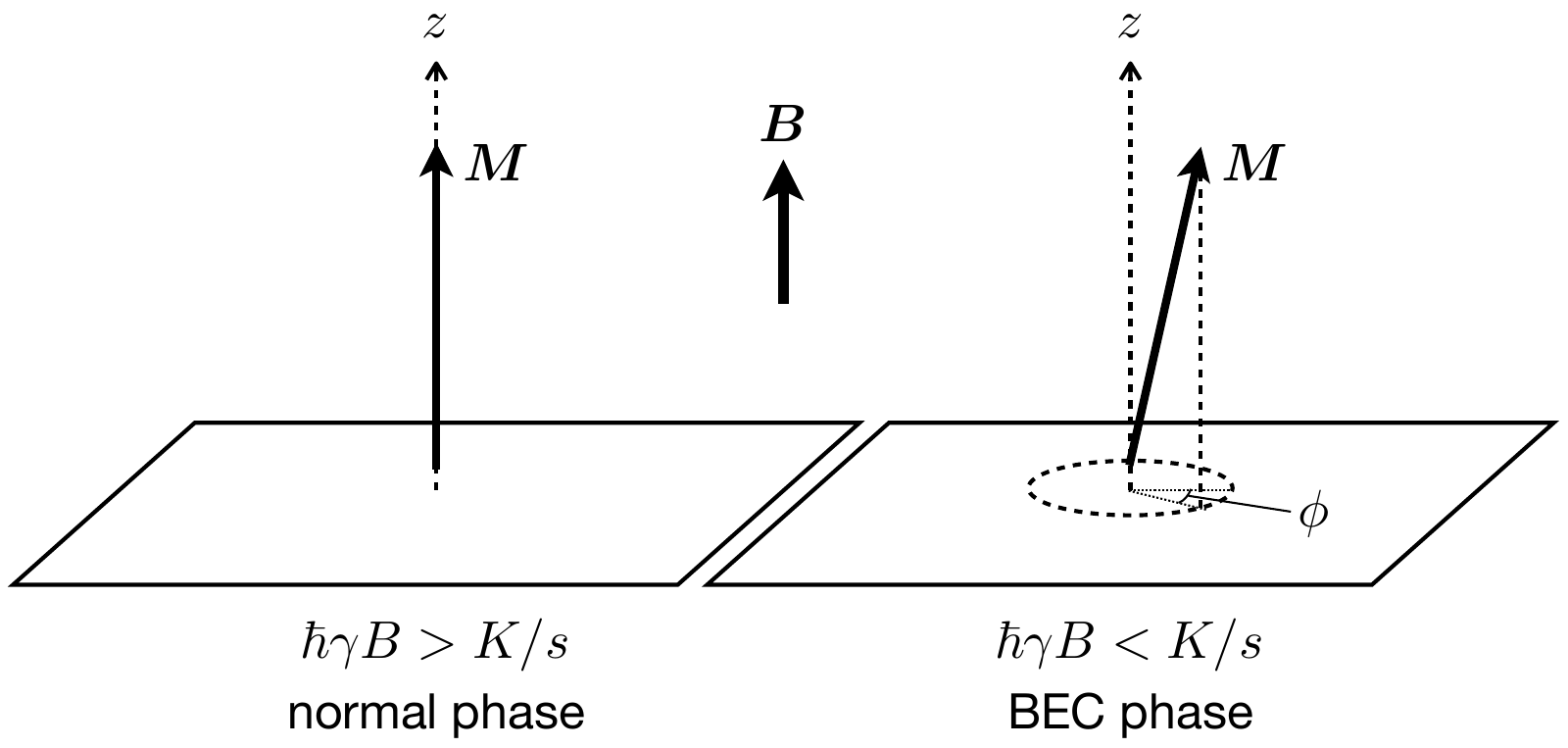}
\caption{Order parameter orientation in the normal and the BEC phases. Canting of the magnetization in the BEC phase produces a planar component of $\bM$ and signals a spontaneous breaking of global U(1) symmetry. The inplane angle $\f$ defines the U(1) phase. }
\label{fig7}
\end{figure}

Equation~\eqref{hf} has global U(1) spin-rotational symmetry that entails conservation of total $s_z$. For large enough fields $B>B_c\equiv K/\hbar s\g$, the system is in the U(1)-symmetric normal phase, where the magnetization points in the $z$ direction, parallel to the field. The system then enters the BEC phase for $B<B_c$, where the order parameter cants away from the $z$ axis. The BEC phase is characterized by a broken U(1) symmetry, where the order parameter's azimuthal angle $\f$ defines the U(1) angle  (see Fig.~\ref{fig7}). 

We focus exclusively on the normal phase, but in the vicinity of the BEC critical point $B\gtrsim B_c$, and perform the Holstein-Primakoff transformation with respect to the ordered moment~\cite{holsteinPR40}. In terms of the magnon operator $a_{\bq k}$, Eq.~\eqref{hf} can be re-expressed as $H_F=H_m+H_{\rm int}$, where $H_m$ describes the free magnons and $H_{\rm int}$ their interactions. The free contribution reads $\smash{H_m=\ve_{\bq k}a^\dag_{\bq k}a_{\bq k}}$, where $\bq$  and $k$ label the inplane and transverse wavevectors, respectively, $\ve_{\bq k}=\De+A(q^2+k^2)/s$ is the magnon spectrum, and $\De=\hbar\g(B-B_c)$ is the magnon gap (see Supplemental Material). We neglect magnon-magnon interactions, as quantum conductivity corrections are typically measured at low temperatures, where the magnons are dilute, so their mutual interactions should not play a crucial role. We account for viscous magnon loss in the FI through phenomenological Gilbert damping.


For a thin magnetic film, we may ignore the variation of the magnetization in the transverse direction and work with the average quantity $\bar\bs(\br)=(1/d)\smallint_{-d}^0dz\,\bs(\br,z)$, where $\bs(\br,z)$ represents the fluctuating portion of the FI spin density. At the interface, the spin density in the metal couples to this average quantity via exchange,
\beq
\label{hc_6}
H_c = -\frac{\cJ v_0}{2}\int d^2\br\,\psi^\dag_{\br\s}\bsig_{\s\s'}\psi_{\br\s'}\cdot\bar\bs(\br)\,,
\eeq
where $\bsig$ is the vector of Pauli matrices, and $v_0$ is the microscopic unit cell volume of the FI. 

\textit{Formalism}:~Within the Kubo formalism, the response function of interest is given by $\Pi^R(\br-\br',t-t')=-(\io/\hbar)\Theta(t-t')\sang{\square{j_c(\br,t),j_c(\br',t')}}$, where $j_c=\sum_\s(e\hbar/2m\io)[\psi_{\br\s}^\dag(\pd_x\psi_{\br\s})-(\pd_x\psi_{\br\s}^\dag)\psi_{\br\s}]$ is the paramagnetic current operator along $x$ (see Fig.~\ref{fig1}). We include the effects of the FI by calculating the response function to $\cJ^2$ order. The conductivity is then produced via $\s=\lim_{\nu\to0}\lim_{\bp\to0} \io\Pi^R(\bp,\nu)/\nu$ after accounting for the diamagnetic component order-by-order in $\cJ$. 

\begin{figure}[t]
\includegraphics[width=0.99\linewidth]{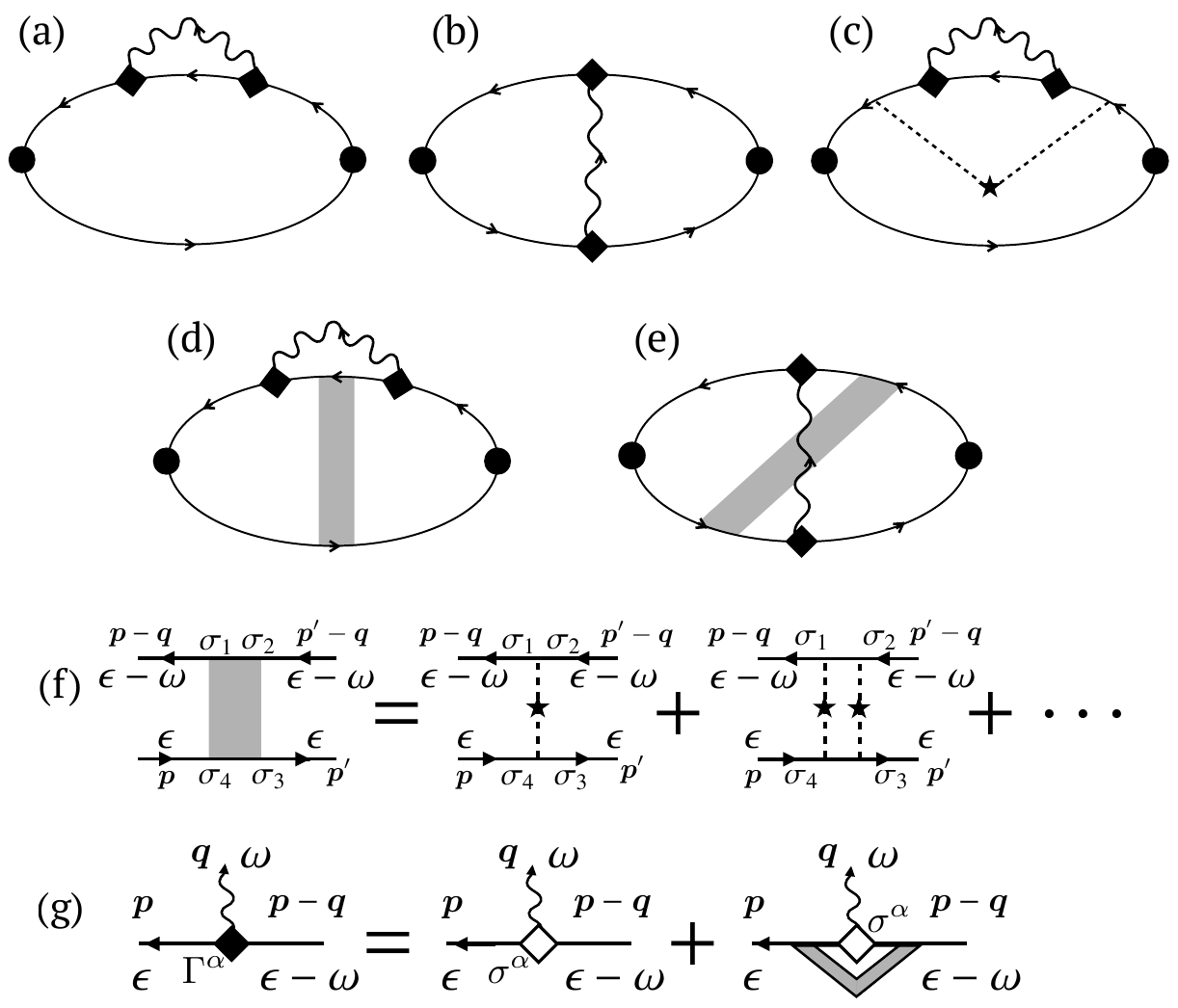}
\caption{(a-e) The contributing conductivity diagrams arising after disorder averaging. Wavy lines represent magnon propagators, shaded grey bars represent impurity ladders, and dashed lines connected by a star represent a single impurity scattering event. (f) A depiction of the ladder function consisting of an infinite sum of non-crossing single impurity scattering events. (g) The dressed spin vertex, depicted in terms of the ladder. The empty diamond denotes the undressed spin vertex $\s^\al$.} 
\label{fig3}
\end{figure}

The current-current correlator to leading order gives the Drude result $\s_D=\s_0k_F\ell$, where $\s_0=e^2/h$. To second-order in $\cJ$ and upon disorder averaging, Figs.~\ref{fig3}(a-e) depicts the leading contributing diagrams. Gray bands represent diffusons (i.e., ladders), the wavy lines the spin fluctuations, and the solid lines denote the electron propagator obtained within the Born approximation, $G^R_{\bk\s}(\w)=(\w-\xi_{\bk\s}/\hbar+\io/2\tau)^{-1}$, where $\xi_{\bk\s}=\hbar^2k^2/2m-\mu+\s\hbar\w_Z/2$ and $\tau$ is the elastic scattering time. The circles represent the usual current vertex.

The Zeeman splitting, $\hbar\w_Z=\hbar\tilde\g B+\cJ S$ ($\tilde\g$ being the gyromagnetic ratio of the metal and $S$ the total spin in the magnetic unit cell), has two contributions: first term due to the external field and the second due to the static magnetization of the FI. This splitting generates a shift in the diffuson pole whenever spins on the upper and lower branches are antiparallel, i.e., the diffuson [depicted in Fig.~\ref{fig3}(f)] reads $(2\p g_0\tau^2\hbar)^{-1}[Dq^2\pm\io(\w+(\s_1-\s_3)\w_Z/2)]^{-1}\de_{\s_1\s_2}\de_{\s_3\s_4}$, where the $+$ ($-$) sign obtains when the propagators on the top side of the ladder are retarded (advanced) and on the bottom side are advanced (retarded), $D=v_F^2\ta/2$ is the diffusion constant, and $g_0=m/2\pi\hbar^2$ is the density of states per spin of the metal. In Figs.~\ref{fig3}(a-e), each bare spin vertex is dressed by this impurity ladder [see Fig.~\ref{fig3}(g)], thus resulting in the renormalized vertex $\G^\al$ (depicted as filled diamonds), where $\al=x,y,z$ labels the spin orientation (see Supplemental Material)~\footnote{The dressed spin vertex is traceless in spin space; therefore, the symmetry of the interaction causes any fermion particle-hole bubble containing only a single interaction, i.e., the so-called Hartree diagrams, to vanish~\cite{syzranovPRL12}. These diagrams have therefore been excluded from Fig.~\ref{fig3}(a-e).}.

{\em Results}:~Diagrams (a),~(b), and~(c) in Fig.~\ref{fig3} cancel exactly. In evaluating the remaining two diagrams, we introduce the FI response function $\chi^R_{\al\be}(\br,t)\equiv-\io\Theta(t)\sang{[\bar s^\al(\br,t),\bar s^\be(\bzero,0)]}$. In the diffusive regime $|\W|\tau,\w_Z\tau\ll1$, $Dq^2\tau\ll1$, the conductivity correction due to spin fluctuations becomes
\begin{multline}
\label{dcmod_6}
\frac{\de\s_s}{\s_0} = -\frac{\pi \cJ^2v_0^2v_F^4\ta^2g_0}{2\hbar}\int\frac{d^2\bq}{(2\pi)^2}\int\frac{d\W}{2\pi}\frac{dF(\W,T)}{d\W}\\
\times \sum_{\s=\pm}{\rm Im}\curly{q^2\frac{\chi^{R}_{-+}(\bq,\W)+\chi^{R}_{+-}(\bq,\W)}{[Dq^2-\io(\W+\s\w_Z)]^3}}\,,
\end{multline}
where the function $F(\W,T)=\W\coth(\hbar\W/2k_BT)$ contains the temperature dependence (see Supplemental Material). The retarded transverse magnon propagator reads (see Supplemental Material) $\chi_{-+}^R(\bq,\W)=(4s/d)(\W-\ve_{\bq0}/\hbar+\io\al|\W|)^{-1}=\chi_{+-}^{R*}(\bq,-\W)$, where $\al$ is the Gilbert damping parameter. 

We first compare Eq.~\eqref{dcmod_6} with the Altshuler-Aronov correction arising from the screened Coulomb interaction. This correction can be obtained by setting $\w_Z=0$ in Eq.~\eqref{dcmod_6} and replacing $\chi^R_{\pm\mp}$ by the propagator for the diffusive charge fluctuations~\cite{altshulerBOOK85,leeRMP85}. In 2D, this leads to an insulating correction $\de\s_c=-c_c\s_0\ln\round{\hbar/k_BT\tau}$, where $c_c$ is an order-1 constant~\cite{altshulerJETP79,altshulerPRL80}.  Unlike the Coulomb case, impurity ladders in Figs.~\ref{fig3}(d,e) connect electrons with opposite spins, which accumulate different phases through the ladder due to the Zeeman field. This generates a shift in the diffuson pole along the frequency axis by $\pm\w_Z$ and renders the $\bq$-integral divergent at $\W=\pm\w_Z$. This divergence occurs due to the long-wavelength sector of the diffuson; at finite temperatures, this infrared divergence can therefore be cut off by the electrons' inelastic dephasing length $\ell_\f$. We implement this cutoff by replacing $Dq^2\rightarrow D(q^2+\ell_\f^{-2})$ in the diffuson.



In evaluating Eq.~\eqref{dcmod_6}, we consider a candidate material suitable for testing our findings: K$_2$CuF$_4$ is an $S=1/2$, quasi-2D, square-lattice FI with an exchange parameter $J=23$~K and easy-plane anisotropy of $0.22$~K~\cite{funahashiSSC76}. A $T\approx0$ BEC transition was achieved in this material at a critical field of $H_c(T\approx0)\approx2.4$~kOe~\cite{hirataPRB17}, and this critical field was found to decrease linearly with temperature, i.e., $H_c(T)=H_c(T\approx0)-aT$, in agreement with the theory of the 2D BEC universality class~\cite{sachdevBOOK11}, and vanish at $T\approx6$~K. 

\begin{figure}[t]
\includegraphics[width=0.87\linewidth]{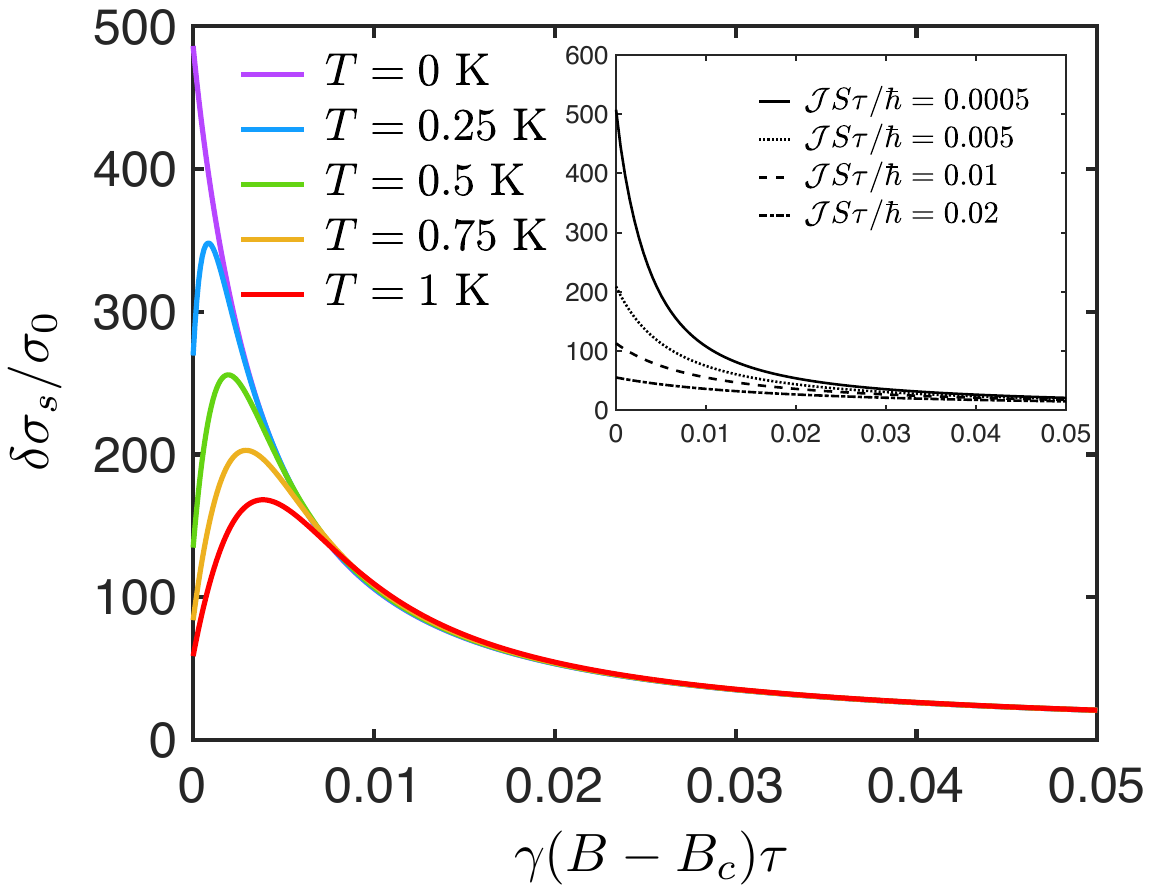}
\caption{(color online) Conductivity correction plotted as a function of the magnon gap $\g(B-B_c)\tau$ for different temperatures and $\cJ S\tau/\hbar=0.0005$; $\de\s_s/\s_0$ is computed in units of $(2S/\p^2)(E_F\tau/\hbar)(\mathcal{J}\tau/\hbar)^2(v_0/d\ell)$. Inset: Conductivity correction for $T=0$ and different exchange biases; $\al=0.1$ is used throughout.} 
\label{fig4}
\end{figure}

In Fig.~\ref{fig4}, Eq.~\eqref{dcmod_6} is plotted as a function of the dimensionless magnon gap $\De\tau/\hbar=\g(B-B_c)\tau$ in units of $(2S/\p^2)(E_F\tau/\hbar)(\mathcal{J}\tau/\hbar)^2(v_0/d\ell^2)$  (see Supplemental Material). Gilbert damping is fixed to $\al=0.1$ throughout, and we fix $\tau=10^{-14}$~s. The conductivity correction is then plotted for temperatures $T=0,0.25,0.5,0.75,1$~K, where the BEC phase is well-defined. A sharp enhancement in the conductivity is obtained as $B\rightarrow B_c$ at both zero and finite temperatures; however, at finite temperatures, the correction is ultimately cutoff by thermal dephasing and converts into a downturn as the magnon gap $\De$ falls below $k_BT$. If we use $\cJ\sim1$~K, K$_2$CuF$_4$ lattice constant of $a\approx5.5\AA$, $d\sim a$, $\ell\sim10$nm, $\ell_\f=10\ell$, and $k_F\ell\sim100$, we have $(2S/\p^2)(E_F\tau/\hbar)(\mathcal{J}\tau/\hbar)^2(v_0/d\ell^2)\sim10^{-7}$. Given the peak values reached in Fig.~\ref{fig4}, we expect a conductivity enhancement of order $\de\s_s/\s_0\sim10^{-5}$ at low temperatures. Since $\s_0=(k_F\ell)(e^2/h)$, this translates to a sheet resistance correction of order $\De R_{\rm sheet}\sim10{\rm m\W}$.

Equation~\eqref{dcmod_6} gives a {\em metallic} correction. Magnetic fluctuations in the FI mediate electron interactions in the triplet channel, which are known to generate positive corrections to the conductivity of disordered electrons~\cite{zalaPRB01,paulPRL05,paulPRB08}. Also, critical magnetic fluctuations here give a non-singular enhancement to the conductivity. This is unlike the problem of disordered electrons coupled to spin fluctuations in the vicinity of a disordered Hertz-Millis transition, where a singular log-squared correction was obtained~\cite{paulPRL05}. This difference arises because the Zeeman gap $\w_Z$ {\em detunes} the diffuson resonances in $(\bq,\W)$-space away from those of the magnon propagator [see Eq.~\eqref{dcmod_6}]. However, an enhancement in the conductivity still ensues as the magnons approach criticality as long as this detuning is weak, i.e., $\w_Z\tau\ll1$. In the inset of Fig.~\ref{fig4}, we verify that the conductivity enhancement vanishes as $\w_Z\tau$ increases.


\textit{Discussion}:~The external field should affect the electrons' orbital motion. The formation of Landau levels due to the perpendicular field should not have any significant effects on our results, as the diffusive condition $\omega_Z\tau\ll1$ implies $\w_c\tau\ll1$ ($\omega_c$ being the cyclotron frequency) and so the levels remain unresolved. Even a weak magnetic field suppresses weak localization, so we neglect localization corrections to the conductivity. Particle-particle scattering can also give rise to a conductivity correction $\de\s_{\rm pp}$. However, for the screened Coulomb interaction, $\de\s_{\rm pp}$ is known to be a smooth function of the field, varying as $\de\s_{\rm pp}\propto h^2$ for $h\ll1$ and $\de\s_{\rm pp}\propto\ln h$ for $h\gg1$, where $h=2DeB/\p k_BT$ in 2D~\cite{altshulerJETP81mr}. We therefore expect no sharp corrections to arise due to the orbital effect as $\De\rightarrow0$ and that it merely leads to a smooth background correction to the sharp enhancement found in this work. 

Many works have studied the BEC of magnetic quasiparticles in dimerized antiferromagnets, which involves a field-induced BEC of spin-1 triplet excitations known as triplons~\cite{nikuniPRL00,giamarchiNATP08,zapfRMP14}. 
A difference between this triplon BEC and the magnon BEC studied in this work is that the triplon band minimum occurs at wavevector $\bQ=(-\p/a,\p/a)$ due to the underlying antiferromagnetic correlations. As a result, spin fluctuations interact strongly with electrons that are close to narrow regions of the Fermi surface, the so-called ‘‘hot spots,’’ which are connected by $\bQ$. This leads to important differences and a separate analysis may be needed, see Refs.~\onlinecite{roschPRL99,syzranovPRL12}. 

There is a body of works investigating {\em nonequilibrium} BEC of magnons, achieved, e.g., by driving a solid film of yttrium-iron garnet (YIG) | a ferrimagnetic insulator | with microwave radiation~\cite{demokritovNAT06}. This nonequilibrium magnon BEC is different from the equilibrium BEC studied here in that the magnon spectrum remains gapped in the former case; nonequilibrium BEC is achieved by raising the magnon chemical potential and is thus attributed to the change in its distribution function. The conductivity corrections computed in this work are sensitive to the magnon spectrum, and the magnon distribution function does not enter Eq.~\eqref{dcmod_6}. Therefore, the same kind of logarithmic correction to the conductivity may not arise at the onset of a nonequilibrium magnon BEC. 

\textit{Conclusion}:~We have studied the conductivity of a disordered metal in which the electrons couple to a magnon system tuned close to a BEC critical point. The combination of metallic disorder and critical magnetic fluctuations, as the magnon gap closes, results in a sharp, though finite, enhancement to the conductivity. The metal-FI bilayer system proposed in this work enables one to detect the onset of a magnon BEC by performing {\em electrical transport measurements} on the adjacent metal. As such, this proposal contributes to a recent body of works explicating how charge transport can be used to probe magnetic insulators in similar metal-insulator bilayers~\cite{nakayamaPRL13,chenPRB13,althammerPRB13,aqeelPRB15,ganzhornPRB16,dongJPCM17,houPRL17,aqeelJPD17,fischerPRB18}. The utility of the proposed bilayer also extends beyond the scope of this work. The possibly of destabilizing a Fermi liquid using magnetic fluctuations offers an exciting arena to engineer unconventional phases. Over the last several years, magnon-induced unconventional superconductivity has been predicted at the surface of topological insulators~\cite{kargarianPRL16} and in metals interfaced by magnetic insulators~\cite{rohlingPRB18,fjaerbuPRB19,brataasPR20}. It would be interesting to explore if any other unconventional phases can be induced in an otherwise trivial metal by coupling it to magnetic fluctuations derived from, e.g., quantum spin liquids or other exotic magnetic phases.

\textit{Acknowledgment}:~We thank V. Galitski and Y. Tserkovnyak for useful discussions. S.~T. acknowledges support by CUNY Research Foundation Project \#90922-07 10 and PSC-CUNY Research Award Program \#63515-00 51.


\begin{thebibliography}{37}%
\makeatletter
\providecommand \@ifxundefined [1]{%
 \@ifx{#1\undefined}
}%
\providecommand \@ifnum [1]{%
 \ifnum #1\expandafter \@firstoftwo
 \else \expandafter \@secondoftwo
 \fi
}%
\providecommand \@ifx [1]{%
 \ifx #1\expandafter \@firstoftwo
 \else \expandafter \@secondoftwo
 \fi
}%
\providecommand \natexlab [1]{#1}%
\providecommand \enquote  [1]{``#1''}%
\providecommand \bibnamefont  [1]{#1}%
\providecommand \bibfnamefont [1]{#1}%
\providecommand \citenamefont [1]{#1}%
\providecommand \href@noop [0]{\@secondoftwo}%
\providecommand \href [0]{\begingroup \@sanitize@url \@href}%
\providecommand \@href[1]{\@@startlink{#1}\@@href}%
\providecommand \@@href[1]{\endgroup#1\@@endlink}%
\providecommand \@sanitize@url [0]{\catcode `\\12\catcode `\$12\catcode
  `\&12\catcode `\#12\catcode `\^12\catcode `\_12\catcode `\%12\relax}%
\providecommand \@@startlink[1]{}%
\providecommand \@@endlink[0]{}%
\providecommand \url  [0]{\begingroup\@sanitize@url \@url }%
\providecommand \@url [1]{\endgroup\@href {#1}{\urlprefix }}%
\providecommand \urlprefix  [0]{URL }%
\providecommand \Eprint [0]{\href }%
\providecommand \doibase [0]{https://doi.org/}%
\providecommand \selectlanguage [0]{\@gobble}%
\providecommand \bibinfo  [0]{\@secondoftwo}%
\providecommand \bibfield  [0]{\@secondoftwo}%
\providecommand \translation [1]{[#1]}%
\providecommand \BibitemOpen [0]{}%
\providecommand \bibitemStop [0]{}%
\providecommand \bibitemNoStop [0]{.\EOS\space}%
\providecommand \EOS [0]{\spacefactor3000\relax}%
\providecommand \BibitemShut  [1]{\csname bibitem#1\endcsname}%
\let\auto@bib@innerbib\@empty
\bibitem [{\citenamefont {Nozi\`eres}\ and\ \citenamefont
  {Pines}(1999)}]{nozieresBOOK99}%
  \BibitemOpen
  \bibfield  {author} {\bibinfo {author} {\bibfnamefont {P.}~\bibnamefont
  {Nozi\`eres}}\ and\ \bibinfo {author} {\bibfnamefont {D.}~\bibnamefont
  {Pines}},\ }\href
  {http://search.ebscohost.com/login.aspx?direct=true&scope=site&db=nlebk&db=nlabk&AN=200517}
  {\emph {\bibinfo {title} {The theory of quantum liquids}}}\ (\bibinfo
  {publisher} {Perseus Books},\ \bibinfo {address} {Cambridge, Mass.},\
  \bibinfo {year} {1999})\BibitemShut {NoStop}%
\bibitem [{\citenamefont {Altshuler}\ and\ \citenamefont
  {Aronov}(1985)}]{altshulerBOOK85}%
  \BibitemOpen
  \bibfield  {author} {\bibinfo {author} {\bibfnamefont {B.}~\bibnamefont
  {Altshuler}}\ and\ \bibinfo {author} {\bibfnamefont {A.}~\bibnamefont
  {Aronov}},\ }in\ \href@noop {} {\emph {\bibinfo {booktitle}
  {Electron--Electron Interactions in Disordered Systems}}},\ \bibinfo {series}
  {Modern Problems in Condensed Matter Sciences}, Vol.~\bibinfo {volume} {10},\
  \bibinfo {editor} {edited by\ \bibinfo {editor} {\bibfnamefont
  {A.}~\bibnamefont {Efros}}\ and\ \bibinfo {editor} {\bibfnamefont
  {M.}~\bibnamefont {Pollak}}}\ (\bibinfo  {publisher} {Elsevier},\ \bibinfo
  {year} {1985})\ pp.\ \bibinfo {pages} {1--153}\BibitemShut {NoStop}%
\bibitem [{\citenamefont {Belitz}\ \emph {et~al.}(2000)\citenamefont {Belitz},
  \citenamefont {Kirkpatrick}, \citenamefont {Narayanan},\ and\ \citenamefont
  {Vojta}}]{belitzPRL00}%
  \BibitemOpen
  \bibfield  {author} {\bibinfo {author} {\bibfnamefont {D.}~\bibnamefont
  {Belitz}}, \bibinfo {author} {\bibfnamefont {T.~R.}\ \bibnamefont
  {Kirkpatrick}}, \bibinfo {author} {\bibfnamefont {R.}~\bibnamefont
  {Narayanan}},\ and\ \bibinfo {author} {\bibfnamefont {T.}~\bibnamefont
  {Vojta}},\ }\href@noop {} {\bibfield  {journal} {\bibinfo  {journal} {Phys.
  Rev. Lett.}\ }\textbf {\bibinfo {volume} {85}},\ \bibinfo {pages} {4602}
  (\bibinfo {year} {2000})}\BibitemShut {NoStop}%
\bibitem [{\citenamefont {Kim}\ and\ \citenamefont {Millis}(2003)}]{kimPRB03}%
  \BibitemOpen
  \bibfield  {author} {\bibinfo {author} {\bibfnamefont {Y.~B.}\ \bibnamefont
  {Kim}}\ and\ \bibinfo {author} {\bibfnamefont {A.~J.}\ \bibnamefont
  {Millis}},\ }\href {https://doi.org/10.1103/PhysRevB.67.085102} {\bibfield
  {journal} {\bibinfo  {journal} {Phys. Rev. B}\ }\textbf {\bibinfo {volume}
  {67}},\ \bibinfo {pages} {085102} (\bibinfo {year} {2003})}\BibitemShut
  {NoStop}%
\bibitem [{\citenamefont {Paul}\ \emph {et~al.}(2005)\citenamefont {Paul},
  \citenamefont {P\'epin}, \citenamefont {Narozhny},\ and\ \citenamefont
  {Maslov}}]{paulPRL05}%
  \BibitemOpen
  \bibfield  {author} {\bibinfo {author} {\bibfnamefont {I.}~\bibnamefont
  {Paul}}, \bibinfo {author} {\bibfnamefont {C.}~\bibnamefont {P\'epin}},
  \bibinfo {author} {\bibfnamefont {B.~N.}\ \bibnamefont {Narozhny}},\ and\
  \bibinfo {author} {\bibfnamefont {D.~L.}\ \bibnamefont {Maslov}},\ }\href
  {https://doi.org/10.1103/PhysRevLett.95.017206} {\bibfield  {journal}
  {\bibinfo  {journal} {Phys. Rev. Lett.}\ }\textbf {\bibinfo {volume} {95}},\
  \bibinfo {pages} {017206} (\bibinfo {year} {2005})}\BibitemShut {NoStop}%
\bibitem [{\citenamefont {Paul}(2008)}]{paulPRB08}%
  \BibitemOpen
  \bibfield  {author} {\bibinfo {author} {\bibfnamefont {I.}~\bibnamefont
  {Paul}},\ }\href@noop {} {\bibfield  {journal} {\bibinfo  {journal} {Phys.
  Rev. B}\ }\textbf {\bibinfo {volume} {77}},\ \bibinfo {pages} {224418}
  (\bibinfo {year} {2008})}\BibitemShut {NoStop}%
\bibitem [{Note1()}]{Note1}%
  \BibitemOpen
  \bibinfo {note} {Non-Fermi liquid temperature scaling has been predicted in
  disordered two-dimensional metals near an antiferromagnet quantum critical
  point as well~\cite {roschPRL99,syzranovPRL12}. This work, however, will
  focus exclusively on critical behavior with a divergent $q=0$
  susceptibility.}\BibitemShut {Stop}%
\bibitem [{\citenamefont {Nikuni}\ \emph {et~al.}(2000)\citenamefont {Nikuni},
  \citenamefont {Oshikawa}, \citenamefont {Oosawa},\ and\ \citenamefont
  {Tanaka}}]{nikuniPRL00}%
  \BibitemOpen
  \bibfield  {author} {\bibinfo {author} {\bibfnamefont {T.}~\bibnamefont
  {Nikuni}}, \bibinfo {author} {\bibfnamefont {M.}~\bibnamefont {Oshikawa}},
  \bibinfo {author} {\bibfnamefont {A.}~\bibnamefont {Oosawa}},\ and\ \bibinfo
  {author} {\bibfnamefont {H.}~\bibnamefont {Tanaka}},\ }\href
  {https://doi.org/10.1103/PhysRevLett.84.5868} {\bibfield  {journal} {\bibinfo
   {journal} {Phys. Rev. Lett.}\ }\textbf {\bibinfo {volume} {84}},\ \bibinfo
  {pages} {5868} (\bibinfo {year} {2000})}\BibitemShut {NoStop}%
\bibitem [{\citenamefont {Rice}(2002)}]{riceSCI02}%
  \BibitemOpen
  \bibfield  {author} {\bibinfo {author} {\bibfnamefont {T.~M.}\ \bibnamefont
  {Rice}},\ }\href {https://doi.org/10.1126/science.1078819} {\bibfield
  {journal} {\bibinfo  {journal} {Science}\ }\textbf {\bibinfo {volume}
  {298}},\ \bibinfo {pages} {760} (\bibinfo {year} {2002})}\BibitemShut
  {NoStop}%
\bibitem [{\citenamefont {Giamarchi}\ \emph {et~al.}(2008)\citenamefont
  {Giamarchi}, \citenamefont {R{\"u}egg},\ and\ \citenamefont
  {Tchernyshyov}}]{giamarchiNATP08}%
  \BibitemOpen
  \bibfield  {author} {\bibinfo {author} {\bibfnamefont {T.}~\bibnamefont
  {Giamarchi}}, \bibinfo {author} {\bibfnamefont {C.}~\bibnamefont
  {R{\"u}egg}},\ and\ \bibinfo {author} {\bibfnamefont {O.}~\bibnamefont
  {Tchernyshyov}},\ }\href@noop {} {\bibfield  {journal} {\bibinfo  {journal}
  {Nature Phys.}\ }\textbf {\bibinfo {volume} {4}},\ \bibinfo {pages} {198}
  (\bibinfo {year} {2008})}\BibitemShut {NoStop}%
\bibitem [{\citenamefont {Zapf}\ \emph {et~al.}(2014)\citenamefont {Zapf},
  \citenamefont {Jaime},\ and\ \citenamefont {Batista}}]{zapfRMP14}%
  \BibitemOpen
  \bibfield  {author} {\bibinfo {author} {\bibfnamefont {V.}~\bibnamefont
  {Zapf}}, \bibinfo {author} {\bibfnamefont {M.}~\bibnamefont {Jaime}},\ and\
  \bibinfo {author} {\bibfnamefont {C.~D.}\ \bibnamefont {Batista}},\
  }\href@noop {} {\bibfield  {journal} {\bibinfo  {journal} {Rev. Mod. Phys.}\
  }\textbf {\bibinfo {volume} {86}},\ \bibinfo {pages} {563} (\bibinfo {year}
  {2014})}\BibitemShut {NoStop}%
\bibitem [{\citenamefont {Holstein}\ and\ \citenamefont
  {Primakoff}(1940)}]{holsteinPR40}%
  \BibitemOpen
  \bibfield  {author} {\bibinfo {author} {\bibfnamefont {T.}~\bibnamefont
  {Holstein}}\ and\ \bibinfo {author} {\bibfnamefont {H.}~\bibnamefont
  {Primakoff}},\ }\href@noop {} {\bibfield  {journal} {\bibinfo  {journal}
  {Phys. Rev.}\ }\textbf {\bibinfo {volume} {58}},\ \bibinfo {pages} {1098}
  (\bibinfo {year} {1940})}\BibitemShut {NoStop}%
\bibitem [{Note2()}]{Note2}%
  \BibitemOpen
  \bibinfo {note} {The dressed spin vertex is traceless in spin space;
  therefore, the symmetry of the interaction causes any fermion particle-hole
  bubble containing only a single interaction, i.e., the so-called Hartree
  diagrams, to vanish~\cite {syzranovPRL12}. These diagrams have therefore been
  excluded from Fig.~\ref {fig3}(a-e).}\BibitemShut {Stop}%
\bibitem [{\citenamefont {Lee}\ and\ \citenamefont
  {Ramakrishnan}(1985)}]{leeRMP85}%
  \BibitemOpen
  \bibfield  {author} {\bibinfo {author} {\bibfnamefont {P.~A.}\ \bibnamefont
  {Lee}}\ and\ \bibinfo {author} {\bibfnamefont {T.~V.}\ \bibnamefont
  {Ramakrishnan}},\ }\href@noop {} {\bibfield  {journal} {\bibinfo  {journal}
  {Rev. Mod. Phys.}\ }\textbf {\bibinfo {volume} {57}},\ \bibinfo {pages} {287}
  (\bibinfo {year} {1985})}\BibitemShut {NoStop}%
\bibitem [{\citenamefont {Altshuler}\ and\ \citenamefont
  {Aronov}(1979)}]{altshulerJETP79}%
  \BibitemOpen
  \bibfield  {author} {\bibinfo {author} {\bibfnamefont {B.~L.}\ \bibnamefont
  {Altshuler}}\ and\ \bibinfo {author} {\bibfnamefont {A.~G.}\ \bibnamefont
  {Aronov}},\ }\href@noop {} {\bibfield  {journal} {\bibinfo  {journal} {Sov.
  Phys. JETP}\ }\textbf {\bibinfo {volume} {50}},\ \bibinfo {pages} {968}
  (\bibinfo {year} {1979})}\BibitemShut {NoStop}%
\bibitem [{\citenamefont {Altshuler}\ \emph {et~al.}(1980)\citenamefont
  {Altshuler}, \citenamefont {Aronov},\ and\ \citenamefont
  {Lee}}]{altshulerPRL80}%
  \BibitemOpen
  \bibfield  {author} {\bibinfo {author} {\bibfnamefont {B.~L.}\ \bibnamefont
  {Altshuler}}, \bibinfo {author} {\bibfnamefont {A.~G.}\ \bibnamefont
  {Aronov}},\ and\ \bibinfo {author} {\bibfnamefont {P.~A.}\ \bibnamefont
  {Lee}},\ }\href {https://doi.org/10.1103/PhysRevLett.44.1288} {\bibfield
  {journal} {\bibinfo  {journal} {Phys. Rev. Lett.}\ }\textbf {\bibinfo
  {volume} {44}},\ \bibinfo {pages} {1288} (\bibinfo {year}
  {1980})}\BibitemShut {NoStop}%
\bibitem [{\citenamefont {Funahashi}\ \emph {et~al.}(1976)\citenamefont
  {Funahashi}, \citenamefont {Moussa},\ and\ \citenamefont
  {Steiner}}]{funahashiSSC76}%
  \BibitemOpen
  \bibfield  {author} {\bibinfo {author} {\bibfnamefont {S.}~\bibnamefont
  {Funahashi}}, \bibinfo {author} {\bibfnamefont {F.}~\bibnamefont {Moussa}},\
  and\ \bibinfo {author} {\bibfnamefont {M.}~\bibnamefont {Steiner}},\
  }\href@noop {} {\bibfield  {journal} {\bibinfo  {journal} {Solid State
  Commun.}\ }\textbf {\bibinfo {volume} {18}},\ \bibinfo {pages} {433}
  (\bibinfo {year} {1976})}\BibitemShut {NoStop}%
\bibitem [{\citenamefont {Hirata}\ \emph {et~al.}(2017)\citenamefont {Hirata},
  \citenamefont {Kurita}, \citenamefont {Yamada},\ and\ \citenamefont
  {Tanaka}}]{hirataPRB17}%
  \BibitemOpen
  \bibfield  {author} {\bibinfo {author} {\bibfnamefont {S.}~\bibnamefont
  {Hirata}}, \bibinfo {author} {\bibfnamefont {N.}~\bibnamefont {Kurita}},
  \bibinfo {author} {\bibfnamefont {M.}~\bibnamefont {Yamada}},\ and\ \bibinfo
  {author} {\bibfnamefont {H.}~\bibnamefont {Tanaka}},\ }\href@noop {}
  {\bibfield  {journal} {\bibinfo  {journal} {Phys. Rev. B}\ }\textbf {\bibinfo
  {volume} {95}},\ \bibinfo {pages} {174406} (\bibinfo {year}
  {2017})}\BibitemShut {NoStop}%
\bibitem [{\citenamefont {Sachdev}(2011)}]{sachdevBOOK11}%
  \BibitemOpen
  \bibfield  {author} {\bibinfo {author} {\bibfnamefont {S.}~\bibnamefont
  {Sachdev}},\ }\href@noop {} {\emph {\bibinfo {title} {Quantum Phase
  Transitions}}},\ \bibinfo {edition} {2nd}\ ed.\ (\bibinfo  {publisher}
  {Cambridge University Press},\ \bibinfo {address} {Cambridge},\ \bibinfo
  {year} {2011})\BibitemShut {NoStop}%
\bibitem [{\citenamefont {Zala}\ \emph {et~al.}(2001)\citenamefont {Zala},
  \citenamefont {Narozhny},\ and\ \citenamefont {Aleiner}}]{zalaPRB01}%
  \BibitemOpen
  \bibfield  {author} {\bibinfo {author} {\bibfnamefont {G.}~\bibnamefont
  {Zala}}, \bibinfo {author} {\bibfnamefont {B.~N.}\ \bibnamefont {Narozhny}},\
  and\ \bibinfo {author} {\bibfnamefont {I.~L.}\ \bibnamefont {Aleiner}},\
  }\href {https://doi.org/10.1103/PhysRevB.64.214204} {\bibfield  {journal}
  {\bibinfo  {journal} {Phys. Rev. B}\ }\textbf {\bibinfo {volume} {64}},\
  \bibinfo {pages} {214204} (\bibinfo {year} {2001})}\BibitemShut {NoStop}%
\bibitem [{\citenamefont {Altshuler}\ \emph {et~al.}(1981)\citenamefont
  {Altshuler}, \citenamefont {Aronov}, \citenamefont {Larkin},\ and\
  \citenamefont {Khmelnitskii}}]{altshulerJETP81mr}%
  \BibitemOpen
  \bibfield  {author} {\bibinfo {author} {\bibfnamefont {B.~L.}\ \bibnamefont
  {Altshuler}}, \bibinfo {author} {\bibfnamefont {A.~G.}\ \bibnamefont
  {Aronov}}, \bibinfo {author} {\bibfnamefont {A.~I.}\ \bibnamefont {Larkin}},\
  and\ \bibinfo {author} {\bibfnamefont {D.~E.}\ \bibnamefont {Khmelnitskii}},\
  }\href@noop {} {\bibfield  {journal} {\bibinfo  {journal} {Sov. Phys. JETP}\
  }\textbf {\bibinfo {volume} {54}},\ \bibinfo {pages} {411} (\bibinfo {year}
  {1981})}\BibitemShut {NoStop}%
\bibitem [{\citenamefont {Rosch}(1999)}]{roschPRL99}%
  \BibitemOpen
  \bibfield  {author} {\bibinfo {author} {\bibfnamefont {A.}~\bibnamefont
  {Rosch}},\ }\href {https://doi.org/10.1103/PhysRevLett.82.4280} {\bibfield
  {journal} {\bibinfo  {journal} {Phys. Rev. Lett.}\ }\textbf {\bibinfo
  {volume} {82}},\ \bibinfo {pages} {4280} (\bibinfo {year}
  {1999})}\BibitemShut {NoStop}%
\bibitem [{\citenamefont {Syzranov}\ and\ \citenamefont
  {Schmalian}(2012)}]{syzranovPRL12}%
  \BibitemOpen
  \bibfield  {author} {\bibinfo {author} {\bibfnamefont {S.~V.}\ \bibnamefont
  {Syzranov}}\ and\ \bibinfo {author} {\bibfnamefont {J.}~\bibnamefont
  {Schmalian}},\ }\href {https://doi.org/10.1103/PhysRevLett.109.156403}
  {\bibfield  {journal} {\bibinfo  {journal} {Phys. Rev. Lett.}\ }\textbf
  {\bibinfo {volume} {109}},\ \bibinfo {pages} {156403} (\bibinfo {year}
  {2012})}\BibitemShut {NoStop}%
\bibitem [{\citenamefont {Demokritov}\ \emph {et~al.}(2006)\citenamefont
  {Demokritov}, \citenamefont {Demidov}, \citenamefont {Dzyapko}, \citenamefont
  {Melkov}, \citenamefont {Serga}, \citenamefont {Hillebrands},\ and\
  \citenamefont {Slavin}}]{demokritovNAT06}%
  \BibitemOpen
  \bibfield  {author} {\bibinfo {author} {\bibfnamefont {S.~O.}\ \bibnamefont
  {Demokritov}}, \bibinfo {author} {\bibfnamefont {V.~E.}\ \bibnamefont
  {Demidov}}, \bibinfo {author} {\bibfnamefont {O.}~\bibnamefont {Dzyapko}},
  \bibinfo {author} {\bibfnamefont {G.~A.}\ \bibnamefont {Melkov}}, \bibinfo
  {author} {\bibfnamefont {A.~A.}\ \bibnamefont {Serga}}, \bibinfo {author}
  {\bibfnamefont {B.}~\bibnamefont {Hillebrands}},\ and\ \bibinfo {author}
  {\bibfnamefont {A.~N.}\ \bibnamefont {Slavin}},\ }\href@noop {} {\bibfield
  {journal} {\bibinfo  {journal} {Nature}\ }\textbf {\bibinfo {volume} {443}},\
  \bibinfo {pages} {430} (\bibinfo {year} {2006})}\BibitemShut {NoStop}%
\bibitem [{\citenamefont {Nakayama}\ \emph {et~al.}(2013)\citenamefont
  {Nakayama}, \citenamefont {Althammer}, \citenamefont {Chen}, \citenamefont
  {Uchida}, \citenamefont {Kajiwara}, \citenamefont {Kikuchi}, \citenamefont
  {Ohtani}, \citenamefont {Gepr\"ags}, \citenamefont {Opel}, \citenamefont
  {Takahashi}, \citenamefont {Gross}, \citenamefont {Bauer}, \citenamefont
  {Goennenwein},\ and\ \citenamefont {Saitoh}}]{nakayamaPRL13}%
  \BibitemOpen
  \bibfield  {author} {\bibinfo {author} {\bibfnamefont {H.}~\bibnamefont
  {Nakayama}}, \bibinfo {author} {\bibfnamefont {M.}~\bibnamefont {Althammer}},
  \bibinfo {author} {\bibfnamefont {Y.-T.}\ \bibnamefont {Chen}}, \bibinfo
  {author} {\bibfnamefont {K.}~\bibnamefont {Uchida}}, \bibinfo {author}
  {\bibfnamefont {Y.}~\bibnamefont {Kajiwara}}, \bibinfo {author}
  {\bibfnamefont {D.}~\bibnamefont {Kikuchi}}, \bibinfo {author} {\bibfnamefont
  {T.}~\bibnamefont {Ohtani}}, \bibinfo {author} {\bibfnamefont
  {S.}~\bibnamefont {Gepr\"ags}}, \bibinfo {author} {\bibfnamefont
  {M.}~\bibnamefont {Opel}}, \bibinfo {author} {\bibfnamefont {S.}~\bibnamefont
  {Takahashi}}, \bibinfo {author} {\bibfnamefont {R.}~\bibnamefont {Gross}},
  \bibinfo {author} {\bibfnamefont {G.~E.~W.}\ \bibnamefont {Bauer}}, \bibinfo
  {author} {\bibfnamefont {S.~T.~B.}\ \bibnamefont {Goennenwein}},\ and\
  \bibinfo {author} {\bibfnamefont {E.}~\bibnamefont {Saitoh}},\ }\href
  {https://doi.org/10.1103/PhysRevLett.110.206601} {\bibfield  {journal}
  {\bibinfo  {journal} {Phys. Rev. Lett.}\ }\textbf {\bibinfo {volume} {110}},\
  \bibinfo {pages} {206601} (\bibinfo {year} {2013})}\BibitemShut {NoStop}%
\bibitem [{\citenamefont {Chen}\ \emph {et~al.}(2013)\citenamefont {Chen},
  \citenamefont {Takahashi}, \citenamefont {Nakayama}, \citenamefont
  {Althammer}, \citenamefont {Goennenwein}, \citenamefont {Saitoh},\ and\
  \citenamefont {Bauer}}]{chenPRB13}%
  \BibitemOpen
  \bibfield  {author} {\bibinfo {author} {\bibfnamefont {Y.-T.}\ \bibnamefont
  {Chen}}, \bibinfo {author} {\bibfnamefont {S.}~\bibnamefont {Takahashi}},
  \bibinfo {author} {\bibfnamefont {H.}~\bibnamefont {Nakayama}}, \bibinfo
  {author} {\bibfnamefont {M.}~\bibnamefont {Althammer}}, \bibinfo {author}
  {\bibfnamefont {S.~T.~B.}\ \bibnamefont {Goennenwein}}, \bibinfo {author}
  {\bibfnamefont {E.}~\bibnamefont {Saitoh}},\ and\ \bibinfo {author}
  {\bibfnamefont {G.~E.~W.}\ \bibnamefont {Bauer}},\ }\href
  {https://doi.org/10.1103/PhysRevB.87.144411} {\bibfield  {journal} {\bibinfo
  {journal} {Phys. Rev. B}\ }\textbf {\bibinfo {volume} {87}},\ \bibinfo
  {pages} {144411} (\bibinfo {year} {2013})}\BibitemShut {NoStop}%
\bibitem [{\citenamefont {Althammer}\ \emph {et~al.}(2013)\citenamefont
  {Althammer}, \citenamefont {Meyer}, \citenamefont {Nakayama}, \citenamefont
  {Schreier}, \citenamefont {Altmannshofer}, \citenamefont {Weiler},
  \citenamefont {Huebl}, \citenamefont {Gepr\"ags}, \citenamefont {Opel},
  \citenamefont {Gross}, \citenamefont {Meier}, \citenamefont {Klewe},
  \citenamefont {Kuschel}, \citenamefont {Schmalhorst}, \citenamefont {Reiss},
  \citenamefont {Shen}, \citenamefont {Gupta}, \citenamefont {Chen},
  \citenamefont {Bauer}, \citenamefont {Saitoh},\ and\ \citenamefont
  {Goennenwein}}]{althammerPRB13}%
  \BibitemOpen
  \bibfield  {author} {\bibinfo {author} {\bibfnamefont {M.}~\bibnamefont
  {Althammer}}, \bibinfo {author} {\bibfnamefont {S.}~\bibnamefont {Meyer}},
  \bibinfo {author} {\bibfnamefont {H.}~\bibnamefont {Nakayama}}, \bibinfo
  {author} {\bibfnamefont {M.}~\bibnamefont {Schreier}}, \bibinfo {author}
  {\bibfnamefont {S.}~\bibnamefont {Altmannshofer}}, \bibinfo {author}
  {\bibfnamefont {M.}~\bibnamefont {Weiler}}, \bibinfo {author} {\bibfnamefont
  {H.}~\bibnamefont {Huebl}}, \bibinfo {author} {\bibfnamefont
  {S.}~\bibnamefont {Gepr\"ags}}, \bibinfo {author} {\bibfnamefont
  {M.}~\bibnamefont {Opel}}, \bibinfo {author} {\bibfnamefont {R.}~\bibnamefont
  {Gross}}, \bibinfo {author} {\bibfnamefont {D.}~\bibnamefont {Meier}},
  \bibinfo {author} {\bibfnamefont {C.}~\bibnamefont {Klewe}}, \bibinfo
  {author} {\bibfnamefont {T.}~\bibnamefont {Kuschel}}, \bibinfo {author}
  {\bibfnamefont {J.-M.}\ \bibnamefont {Schmalhorst}}, \bibinfo {author}
  {\bibfnamefont {G.}~\bibnamefont {Reiss}}, \bibinfo {author} {\bibfnamefont
  {L.}~\bibnamefont {Shen}}, \bibinfo {author} {\bibfnamefont {A.}~\bibnamefont
  {Gupta}}, \bibinfo {author} {\bibfnamefont {Y.-T.}\ \bibnamefont {Chen}},
  \bibinfo {author} {\bibfnamefont {G.~E.~W.}\ \bibnamefont {Bauer}}, \bibinfo
  {author} {\bibfnamefont {E.}~\bibnamefont {Saitoh}},\ and\ \bibinfo {author}
  {\bibfnamefont {S.~T.~B.}\ \bibnamefont {Goennenwein}},\ }\href
  {https://doi.org/10.1103/PhysRevB.87.224401} {\bibfield  {journal} {\bibinfo
  {journal} {Phys. Rev. B}\ }\textbf {\bibinfo {volume} {87}},\ \bibinfo
  {pages} {224401} (\bibinfo {year} {2013})}\BibitemShut {NoStop}%
\bibitem [{\citenamefont {Aqeel}\ \emph {et~al.}(2015)\citenamefont {Aqeel},
  \citenamefont {Vlietstra}, \citenamefont {Heuver}, \citenamefont {Bauer},
  \citenamefont {Noheda}, \citenamefont {van Wees},\ and\ \citenamefont
  {Palstra}}]{aqeelPRB15}%
  \BibitemOpen
  \bibfield  {author} {\bibinfo {author} {\bibfnamefont {A.}~\bibnamefont
  {Aqeel}}, \bibinfo {author} {\bibfnamefont {N.}~\bibnamefont {Vlietstra}},
  \bibinfo {author} {\bibfnamefont {J.~A.}\ \bibnamefont {Heuver}}, \bibinfo
  {author} {\bibfnamefont {G.~E.~W.}\ \bibnamefont {Bauer}}, \bibinfo {author}
  {\bibfnamefont {B.}~\bibnamefont {Noheda}}, \bibinfo {author} {\bibfnamefont
  {B.~J.}\ \bibnamefont {van Wees}},\ and\ \bibinfo {author} {\bibfnamefont
  {T.~T.~M.}\ \bibnamefont {Palstra}},\ }\href
  {https://doi.org/10.1103/PhysRevB.92.224410} {\bibfield  {journal} {\bibinfo
  {journal} {Phys. Rev. B}\ }\textbf {\bibinfo {volume} {92}},\ \bibinfo
  {pages} {224410} (\bibinfo {year} {2015})}\BibitemShut {NoStop}%
\bibitem [{\citenamefont {Ganzhorn}\ \emph {et~al.}(2016)\citenamefont
  {Ganzhorn}, \citenamefont {Barker}, \citenamefont {Schlitz}, \citenamefont
  {Piot}, \citenamefont {Ollefs}, \citenamefont {Guillou}, \citenamefont
  {Wilhelm}, \citenamefont {Rogalev}, \citenamefont {Opel}, \citenamefont
  {Althammer}, \citenamefont {Gepr\"ags}, \citenamefont {Huebl}, \citenamefont
  {Gross}, \citenamefont {Bauer},\ and\ \citenamefont
  {Goennenwein}}]{ganzhornPRB16}%
  \BibitemOpen
  \bibfield  {author} {\bibinfo {author} {\bibfnamefont {K.}~\bibnamefont
  {Ganzhorn}}, \bibinfo {author} {\bibfnamefont {J.}~\bibnamefont {Barker}},
  \bibinfo {author} {\bibfnamefont {R.}~\bibnamefont {Schlitz}}, \bibinfo
  {author} {\bibfnamefont {B.~A.}\ \bibnamefont {Piot}}, \bibinfo {author}
  {\bibfnamefont {K.}~\bibnamefont {Ollefs}}, \bibinfo {author} {\bibfnamefont
  {F.}~\bibnamefont {Guillou}}, \bibinfo {author} {\bibfnamefont
  {F.}~\bibnamefont {Wilhelm}}, \bibinfo {author} {\bibfnamefont
  {A.}~\bibnamefont {Rogalev}}, \bibinfo {author} {\bibfnamefont
  {M.}~\bibnamefont {Opel}}, \bibinfo {author} {\bibfnamefont {M.}~\bibnamefont
  {Althammer}}, \bibinfo {author} {\bibfnamefont {S.}~\bibnamefont
  {Gepr\"ags}}, \bibinfo {author} {\bibfnamefont {H.}~\bibnamefont {Huebl}},
  \bibinfo {author} {\bibfnamefont {R.}~\bibnamefont {Gross}}, \bibinfo
  {author} {\bibfnamefont {G.~E.~W.}\ \bibnamefont {Bauer}},\ and\ \bibinfo
  {author} {\bibfnamefont {S.~T.~B.}\ \bibnamefont {Goennenwein}},\ }\href
  {https://doi.org/10.1103/PhysRevB.94.094401} {\bibfield  {journal} {\bibinfo
  {journal} {Phys. Rev. B}\ }\textbf {\bibinfo {volume} {94}},\ \bibinfo
  {pages} {094401} (\bibinfo {year} {2016})}\BibitemShut {NoStop}%
\bibitem [{\citenamefont {Dong}\ \emph {et~al.}(2017)\citenamefont {Dong},
  \citenamefont {Cramer}, \citenamefont {Ganzhorn}, \citenamefont {Yuan},
  \citenamefont {Guo}, \citenamefont {Goennenwein},\ and\ \citenamefont
  {Kl{\"a}ui}}]{dongJPCM17}%
  \BibitemOpen
  \bibfield  {author} {\bibinfo {author} {\bibfnamefont {B.-W.}\ \bibnamefont
  {Dong}}, \bibinfo {author} {\bibfnamefont {J.}~\bibnamefont {Cramer}},
  \bibinfo {author} {\bibfnamefont {K.}~\bibnamefont {Ganzhorn}}, \bibinfo
  {author} {\bibfnamefont {H.~Y.}\ \bibnamefont {Yuan}}, \bibinfo {author}
  {\bibfnamefont {E.-J.}\ \bibnamefont {Guo}}, \bibinfo {author} {\bibfnamefont
  {S.~T.~B.}\ \bibnamefont {Goennenwein}},\ and\ \bibinfo {author}
  {\bibfnamefont {M.}~\bibnamefont {Kl{\"a}ui}},\ }\href
  {https://doi.org/10.1088/1361-648x/aa9e26} {\bibfield  {journal} {\bibinfo
  {journal} {Journal of Physics: Condensed Matter}\ }\textbf {\bibinfo {volume}
  {30}},\ \bibinfo {pages} {035802} (\bibinfo {year} {2017})}\BibitemShut
  {NoStop}%
\bibitem [{\citenamefont {Hou}\ \emph {et~al.}(2017)\citenamefont {Hou},
  \citenamefont {Qiu}, \citenamefont {Barker}, \citenamefont {Sato},
  \citenamefont {Yamamoto}, \citenamefont {V\'elez}, \citenamefont
  {Gomez-Perez}, \citenamefont {Hueso}, \citenamefont {Casanova},\ and\
  \citenamefont {Saitoh}}]{houPRL17}%
  \BibitemOpen
  \bibfield  {author} {\bibinfo {author} {\bibfnamefont {D.}~\bibnamefont
  {Hou}}, \bibinfo {author} {\bibfnamefont {Z.}~\bibnamefont {Qiu}}, \bibinfo
  {author} {\bibfnamefont {J.}~\bibnamefont {Barker}}, \bibinfo {author}
  {\bibfnamefont {K.}~\bibnamefont {Sato}}, \bibinfo {author} {\bibfnamefont
  {K.}~\bibnamefont {Yamamoto}}, \bibinfo {author} {\bibfnamefont
  {S.}~\bibnamefont {V\'elez}}, \bibinfo {author} {\bibfnamefont {J.~M.}\
  \bibnamefont {Gomez-Perez}}, \bibinfo {author} {\bibfnamefont {L.~E.}\
  \bibnamefont {Hueso}}, \bibinfo {author} {\bibfnamefont {F.}~\bibnamefont
  {Casanova}},\ and\ \bibinfo {author} {\bibfnamefont {E.}~\bibnamefont
  {Saitoh}},\ }\href {https://doi.org/10.1103/PhysRevLett.118.147202}
  {\bibfield  {journal} {\bibinfo  {journal} {Phys. Rev. Lett.}\ }\textbf
  {\bibinfo {volume} {118}},\ \bibinfo {pages} {147202} (\bibinfo {year}
  {2017})}\BibitemShut {NoStop}%
\bibitem [{\citenamefont {Aqeel}\ \emph {et~al.}(2017)\citenamefont {Aqeel},
  \citenamefont {Mostovoy}, \citenamefont {van Wees},\ and\ \citenamefont
  {Palstra}}]{aqeelJPD17}%
  \BibitemOpen
  \bibfield  {author} {\bibinfo {author} {\bibfnamefont {A.}~\bibnamefont
  {Aqeel}}, \bibinfo {author} {\bibfnamefont {M.}~\bibnamefont {Mostovoy}},
  \bibinfo {author} {\bibfnamefont {B.~J.}\ \bibnamefont {van Wees}},\ and\
  \bibinfo {author} {\bibfnamefont {T.~T.~M.}\ \bibnamefont {Palstra}},\ }\href
  {https://doi.org/10.1088/1361-6463/aa6670} {\bibfield  {journal} {\bibinfo
  {journal} {J. Phys. D: Appl. Phys.}\ }\textbf {\bibinfo {volume} {50}},\
  \bibinfo {pages} {174006} (\bibinfo {year} {2017})}\BibitemShut {NoStop}%
\bibitem [{\citenamefont {Fischer}\ \emph {et~al.}(2018)\citenamefont
  {Fischer}, \citenamefont {Gomonay}, \citenamefont {Schlitz}, \citenamefont
  {Ganzhorn}, \citenamefont {Vlietstra}, \citenamefont {Althammer},
  \citenamefont {Huebl}, \citenamefont {Opel}, \citenamefont {Gross},
  \citenamefont {Goennenwein},\ and\ \citenamefont {Gepr\"ags}}]{fischerPRB18}%
  \BibitemOpen
  \bibfield  {author} {\bibinfo {author} {\bibfnamefont {J.}~\bibnamefont
  {Fischer}}, \bibinfo {author} {\bibfnamefont {O.}~\bibnamefont {Gomonay}},
  \bibinfo {author} {\bibfnamefont {R.}~\bibnamefont {Schlitz}}, \bibinfo
  {author} {\bibfnamefont {K.}~\bibnamefont {Ganzhorn}}, \bibinfo {author}
  {\bibfnamefont {N.}~\bibnamefont {Vlietstra}}, \bibinfo {author}
  {\bibfnamefont {M.}~\bibnamefont {Althammer}}, \bibinfo {author}
  {\bibfnamefont {H.}~\bibnamefont {Huebl}}, \bibinfo {author} {\bibfnamefont
  {M.}~\bibnamefont {Opel}}, \bibinfo {author} {\bibfnamefont {R.}~\bibnamefont
  {Gross}}, \bibinfo {author} {\bibfnamefont {S.~T.~B.}\ \bibnamefont
  {Goennenwein}},\ and\ \bibinfo {author} {\bibfnamefont {S.}~\bibnamefont
  {Gepr\"ags}},\ }\href {https://doi.org/10.1103/PhysRevB.97.014417} {\bibfield
   {journal} {\bibinfo  {journal} {Phys. Rev. B}\ }\textbf {\bibinfo {volume}
  {97}},\ \bibinfo {pages} {014417} (\bibinfo {year} {2018})}\BibitemShut
  {NoStop}%
\bibitem [{\citenamefont {Kargarian}\ \emph {et~al.}(2016)\citenamefont
  {Kargarian}, \citenamefont {Efimkin},\ and\ \citenamefont
  {Galitski}}]{kargarianPRL16}%
  \BibitemOpen
  \bibfield  {author} {\bibinfo {author} {\bibfnamefont {M.}~\bibnamefont
  {Kargarian}}, \bibinfo {author} {\bibfnamefont {D.~K.}\ \bibnamefont
  {Efimkin}},\ and\ \bibinfo {author} {\bibfnamefont {V.}~\bibnamefont
  {Galitski}},\ }\href@noop {} {\bibfield  {journal} {\bibinfo  {journal}
  {Phys. Rev. Lett.}\ }\textbf {\bibinfo {volume} {117}},\ \bibinfo {pages}
  {076806} (\bibinfo {year} {2016})}\BibitemShut {NoStop}%
\bibitem [{\citenamefont {Rohling}\ \emph {et~al.}(2018)\citenamefont
  {Rohling}, \citenamefont {Fj\ae{}rbu},\ and\ \citenamefont
  {Brataas}}]{rohlingPRB18}%
  \BibitemOpen
  \bibfield  {author} {\bibinfo {author} {\bibfnamefont {N.}~\bibnamefont
  {Rohling}}, \bibinfo {author} {\bibfnamefont {E.~L.}\ \bibnamefont
  {Fj\ae{}rbu}},\ and\ \bibinfo {author} {\bibfnamefont {A.}~\bibnamefont
  {Brataas}},\ }\href@noop {} {\bibfield  {journal} {\bibinfo  {journal} {Phys.
  Rev. B}\ }\textbf {\bibinfo {volume} {97}},\ \bibinfo {pages} {115401}
  (\bibinfo {year} {2018})}\BibitemShut {NoStop}%
\bibitem [{\citenamefont {Fj\ae{}rbu}\ \emph {et~al.}(2019)\citenamefont
  {Fj\ae{}rbu}, \citenamefont {Rohling},\ and\ \citenamefont
  {Brataas}}]{fjaerbuPRB19}%
  \BibitemOpen
  \bibfield  {author} {\bibinfo {author} {\bibfnamefont {E.~L.}\ \bibnamefont
  {Fj\ae{}rbu}}, \bibinfo {author} {\bibfnamefont {N.}~\bibnamefont
  {Rohling}},\ and\ \bibinfo {author} {\bibfnamefont {A.}~\bibnamefont
  {Brataas}},\ }\href@noop {} {\bibfield  {journal} {\bibinfo  {journal} {Phys.
  Rev. B}\ }\textbf {\bibinfo {volume} {100}},\ \bibinfo {pages} {125432}
  (\bibinfo {year} {2019})}\BibitemShut {NoStop}%
\bibitem [{\citenamefont {Brataas}\ \emph {et~al.}(2020)\citenamefont
  {Brataas}, \citenamefont {{van Wees}}, \citenamefont {Klein}, \citenamefont
  {{de Loubens}},\ and\ \citenamefont {Viret}}]{brataasPR20}%
  \BibitemOpen
  \bibfield  {author} {\bibinfo {author} {\bibfnamefont {A.}~\bibnamefont
  {Brataas}}, \bibinfo {author} {\bibfnamefont {B.}~\bibnamefont {{van Wees}}},
  \bibinfo {author} {\bibfnamefont {O.}~\bibnamefont {Klein}}, \bibinfo
  {author} {\bibfnamefont {G.}~\bibnamefont {{de Loubens}}},\ and\ \bibinfo
  {author} {\bibfnamefont {M.}~\bibnamefont {Viret}},\ }\href
  {https://doi.org/https://doi.org/10.1016/j.physrep.2020.08.006} {\bibfield
  {journal} {\bibinfo  {journal} {Phys. Rep.}\ }\textbf {\bibinfo {volume}
  {885}},\ \bibinfo {pages} {1} (\bibinfo {year} {2020})}\BibitemShut {NoStop}%
\end{thebibliography}

%

\end{document}